\documentclass[prb,notitlepage,twocolumn,preprintnumbers,floatfix]{revtex4-1}

\usepackage{graphics}

\usepackage{amsthm,amsmath,amssymb,amsfonts,bm}

\usepackage{graphicx}
\usepackage[caption=false]{subfig}

\newcommand{\bx} {{\bf x}}

\newcommand{\bz} {{\bf z}}

\newcommand{\ba} {{\bf a}}
\newcommand{\Q} {{\bf Q}}

\newcommand{\Diag}[1] {\mbox{Diag}\left( #1 \right)}

\begin{document}

\title{Phase retrieval by power iterations}

\begin{abstract}
  I show that the power iteration method applied to the phase
  retrieval problem converges under special conditions.
  One is given the relative phases between small non-overlapping
  groups of pixels of a recorded intensity pattern, but no information
  on the phase between the groups of pixels. Numerical tests show
  that the inverse block iteration recovers the solution in 1
  iteration.
\end{abstract}

\preprint{LBNL-5939E}
\author{Stefano Marchesini }
\affiliation{Advanced Light Source, Lawrence Berkeley National Laboratory, Berkeley, CA 94720}

\maketitle

\section{Introduction}

Given a set of intensity measurements, $a^2\in \mathbb{R}^M$,
an unknown object $\psi\in \mathbb{C}^N$ represented by a
complex $n\times n$ image ($N=n^2$), a known ``illumination matrix'' or
support matrix $\Q$ ($\mathbb{C}^{M\times N}$ matrix), a known propagation operator
(typically one, or a stack of 2D FFT operators) $\bm F$ of dimension $M\times M$ and set
of frames $\bz \in \mathbb{C}^M$, which are related  by:
$$
 \bz=\bm F \Q\psi,\qquad |\bz|=a.
$$
Our goal is to find $\psi$ or the intermediate variable $\bz$, given $\bm F$, $\Q$ and $a$. To do so, we need to find a phase $\phi$
such that  $\bz=a \phi$ is in the range of $\bm F \Q$.

We can eliminate $\psi$ by using the operator $P_\Q
$  to project a vector $\bz$ onto the range of $\bm F \Q$:
\begin{eqnarray}
\label{projections}
P_\Q &=& \bm F \Q (\Q^{\ast}\Q)^{-1}\Q^{\ast}\bm F^\ast. 
\end{eqnarray}
Which ensures that the unknown vector $\psi$ can be obtained from the
frame $\bz$ by $\psi=(\Q^\ast\Q)^{-1}\Q^\ast \bm F^\ast \bz$.

%this is also described in Fig. \ref{fig:exp}.
%\begin{figure}[htbp]
%    \includegraphics[width=0.5\textwidth]{expsetup.pdf}
% \caption{(Left) Experimental geometry in ptychography: An unknown
%   sample with transmission $\psi(\bm r)$ is rastered through an
%   illuminating beam $w(\bm r)$, and a sequence of diffraction
%   measurements $|a_{(i)}|^2$ is recorded on an area detector as the
%   sample is rastered around. The point-wise product between illuminating function and sample $z_{(i)}(\bm r)=w(\br)\psi(\br+\bx_{(i)})$ is related
%to the measurement by a Fourier magnitude relationship 
%$a_{(i)}(\bq) =
%\left |{\cal F} z_{(i)} \right |$.
%\label{fig:exp}}
%\end{figure}
\noindent
A popular approach is to find a vector $z$ such that:
\begin{eqnarray}
\label{eq:erq}
\left \|
[I-P_Q] \bz
\right \|^2&=&0,\\
\label{eq:erf}
\left \|
[I-P_a] \bz
\right \|^2&=&\||\bz|-a\|^2=0,
\end{eqnarray}
are satisfied simultaneously, and where the \textit{Fourier magnitude} projection $P_a$ when applied to a vector $\bz$, yields:
\begin{eqnarray}
\label{projection_f}
P_{\ba} \bz =  \bm{a} \frac{\bz}{| \bz|},\quad \bm{a}=\Diag{a} \ \
\end{eqnarray}
where division are intended as element-wise operations
\cite{augmented,Marchesini2007a}.

\section{Phase optimization}
Here we want to minimize Eq. \ref{eq:erq} w.r.t. a phase vector $\phi$
($\phi_i^\ast \phi_i=1, \forall i$).
That is, we want to find:
\begin{eqnarray}
\nonumber
\label{eq:phase1}
\arg \min_\phi &\quad& \left \| [I - P_\Q ] \mathrm{Diag} (a) \phi \right \|^2,\\
\label{eq:phase2}
\arg \min_\phi        &\quad&\phi^\ast \bm a\left  [ I-P_\Q \right ] \bm a \phi, 
\end{eqnarray}
I discuss three approaches that relax the phase modulus condition
($\phi_i^\ast\phi_i=1\quad \forall i$) to synchronize the relative
phases.
\paragraph{Power iteration}
By changing variable
$\bz=\bm a \phi$, we write:
\begin{eqnarray}
\label{eq:synchronization}
\arg \min_\bz \quad \bz^\ast\left (I- P_\Q \right )\bz
\end{eqnarray}
By relaxing $(a_i^2|\phi_i|^2=a_i^2, \forall i)$ and using $\|\bz\|=\|\bm
a\phi\|^2=\|\bm a\|^2$, we can re-write Eq. (\ref{eq:synchronization})
as finding the eigenvector
with largest eigenvalue\cite{singer}. 
Since $\|\bz\|=\|\bm a \phi\|=\|a\|$ is constant,  we rewrite Eq, (\ref{eq:phase2}) as:
\begin{eqnarray}
\label{eq:phase3a}
\arg \max_\bz \quad \bz^\ast P_\Q \bz,\quad
\end{eqnarray}
we apply one step of power iteration: 
\begin{eqnarray}
\label{eq:phase3}
\nu^{\ell+1}= P_\Q \bz
\end{eqnarray}
We then form a projection on the unit torus to ensure that $|\nu^{\ell+1}_i|=1$ (or $|\bz_i|=a_i$) by element-wise normalization:
\begin{eqnarray}
\nonumber
z^{(\ell+1)}=\bm a \frac{\nu}{|\nu|}&=&P_{a}P_\Q z^{(\ell)}
\end{eqnarray}
Here we have obtained the classical alternating projection method,
which is known to stagnate with classical CDI but to converge (slowly)
in ptychographic imaging.
\paragraph{Greedy phase optimization}
\noindent
Since the diagonal term $\bz^\ast \Diag{P_\Q} \bz$ is also independent
on the choice of $\phi$ (for $|\phi_i|=1$), one can remove it 
 when computing the power iteration:
\begin{eqnarray}
\label{eq:phasecut}
\nu^{(\ell+1)}= [ P_Q -\Diag{P_Q}]\bz^{(\ell)}
\end{eqnarray}
After we apply the projection of $\phi^{\ell+1}$ to the unit torus, we obtain
the following update \cite{Waldspurger}:
$$
\label{eq:phasecut1}
z^{(\ell+1)}=P_a\left (P_\Q  -\Diag{P_Q} \right )z^{(\ell)}
$$
In classical CDI, $P_{Q_{ii}}={\|\Q\|^2\over\|\bm 1\|^2}$ is simply
the sum of the support volume (or area) normalized by the oversampled
volume, in ptychographic imaging $\Diag{P_\Q}$ is the ratio of
intensities $P_{Q_{ii}}={\|Q_i\|^2\over{\|\Q\|^2}}$ for every pixel of
a frame $i$ generate by a submatrix $Q_{i}$.  At the first
iteration, using data generated from the object in Fig. \ref{fig:psi}
with a random phase as a starting guess, Eq. (\ref{eq:phasecut})
appears to out-perform Eq.  (\ref{eq:synchronization}), however the
two methods converge to similar local minimum within ten iterations.
The relaxations in Eqs. (\ref{eq:synchronization},\ref{eq:phasecut})
are similar. By removing diagonal components we change the relaxation.
In Eq. (\ref{eq:synchronization}) we have $\|\bm a^2 \phi\|$ constant,
in Eq. (\ref{eq:phasecut}) $\|[1-\Diag{P_\Q}]\bm a^2 \phi\|$ is
constant. However $\Diag{P_Q}$ is often constant and the two
relaxations are equivalent, giving more weight to high intensity
values.

\paragraph{Inverse iteration\cite{augmented}.}
If we solve the minimization problem 
(Eq.  (\ref{eq:phase2}) 
with a different relaxation,  setting $\|\phi \|=k$ to a constant, 
we re-write the problem as
\begin{eqnarray}
\arg \max \phi^\ast H^{-1 }\phi,
\quad H=\ba [I -P]\ba
\end{eqnarray}
and apply the power iteration:
\begin{eqnarray}
\label{eq:augmented}
H
\nu^{(\ell+1)}= 
\phi^{\ell}
\end{eqnarray}
This method is commonly referred to as inverse iteration and it is
used to find the smallest eigenvector of a matrix. We note however
that any $\nu$ written in the following way:
\begin{equation}
\label{eq:oblique}
\nu=\left (\ba^{-1}  P_\Q \ba\right ) \bx
\end{equation}
is an eigenvector with 0 eigenvalue of $\ba (I-P_\Q) \ba$, therefore
the inverse iteration method cannot be applied directly. 

When $H$ is singular, then instead of power iteration we may want to
find the smallest modification of the phase that is in the null space
of $H$, which we can write it as a
re-weighted LSQ problem of the form:
\begin{eqnarray}
\label{eq:reweight}
\arg \min_\bz \left \|\frac{1}{\ba} (\bz-\bz^\ell) \right \|,
\quad \text{s.t. $\bz= P_\Q \bz$}
\end{eqnarray}
which provides a search direction toward the solution that differs
from standard projection algorithms.  Another approach is to include
additional restrictions on $\nu$ before applying the inverse iteration as described in the following.

\paragraph{Inverse block iteration}
In \cite{augmented} it was observed that computing the exact solution
to Eq. (\ref{eq:augmented}) after ``binning'', or fixing the
relative phase between groups of pixels,  improved
convergence rate in large scale ptychographic imaging.  The use
eigensolvers for the interferometric case was also suggested in
\cite{bandeira}, for the connection Laplacian of a graph.

Let us introduce a binning
matrix $\bm T^\ast$ composed of a series of masks $T_i$ that integrate
over a region of dimension $M/k$ of the data (in Fourier domain). For
example, we can partition our data in 3, creating a tall matrix of
dimension $M\times 3$:
$$
\bm T=\left (\begin{array}{c}
{\bm 1_{M/3}, \bm 0_{M/3},\bm 0_{M/3}}\\
{\bm 0_{M/3}, \bm 1_{M/3},\bm 0_{M/3}}\\
{\bm 0_{M/3}, \bm 0_{M/3},\bm 1_{M/3}}
\end{array}\right )
$$
where $T_1=(\bm 1^\ast_{M/3},\bm 0^\ast_{M/3},\bm 0^\ast_{M/3})^\ast$ is a
vector of length $M$.

We restrict our search of the solution to  Eq. (\ref{eq:augmented})
by restricting $\nu$ to be:
\begin{eqnarray}
\label{eq:augmented1}
\nu=\Diag{\phi^{\ell}} T \omega,
\end{eqnarray}

If we multiply from the left by $T^\ast \Diag{\phi^\ell}^\ast$ in Eq.
(\ref{eq:augmented}) we obtain the inverse iteration step with 
initial 0-phase vector as first guess:
\begin{eqnarray}
\label{eq:augmented2}
\hat H^{(\ell)} \omega^{(\ell+1)}=\lambda_1 \bm 1, \quad 
\end{eqnarray}
Where   
$$
\hat H^{(\ell)}=
T^\ast \Diag{z^{(\ell)}}^\ast [I-P_\Q]\Diag{ z^{(\ell)}} T,
$$
 $\lambda_1$ is a scalar multiplicative factor, and $\bm 1$ is a vector of appropriate length ($=3$ in this example).
By computing $\omega^{(\ell+1)}$ from Eq. (\ref{eq:augmented2}), and $\nu^{\ell+1}$ from Eq. (\ref{eq:augmented1}),
 and projecting on the unit torus we obtain the update $\phi^{(\ell+1)}$:
\begin{eqnarray}
\label{eq:augmented3}
\phi^{(\ell+1)}=\frac{\nu^{(\ell+1)}}{|\nu^{(\ell+1)}|}=\Diag{\phi^{(\ell)}} \frac{T \omega^{(\ell+1)}}{|T \omega^{(\ell+1)}|}.
\end{eqnarray}

In the following section we'll show an example of the inverse iteration method.

\section{Numerical example}
\begin{figure*}[htb]
\begin{minipage}[p]{0.45\linewidth}
\centering
    \includegraphics[width=1\linewidth]{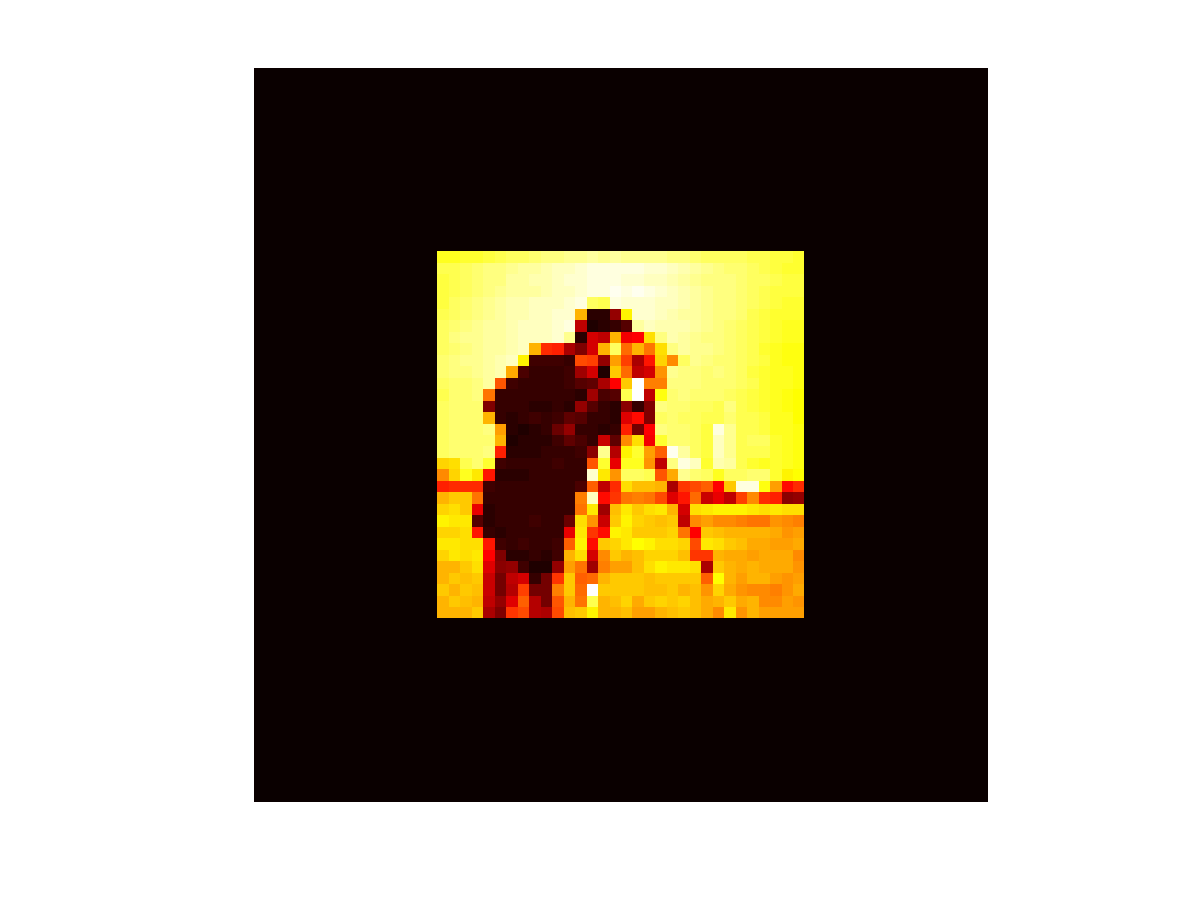}
\caption{\label{fig:psi} Object $\psi$ ($64\times 64$) used to simulate diffraction data}
\end{minipage}
\hspace{0.05\linewidth}
\begin{minipage}[p]{0.45\linewidth}
\centering
    \includegraphics[width=1\linewidth]{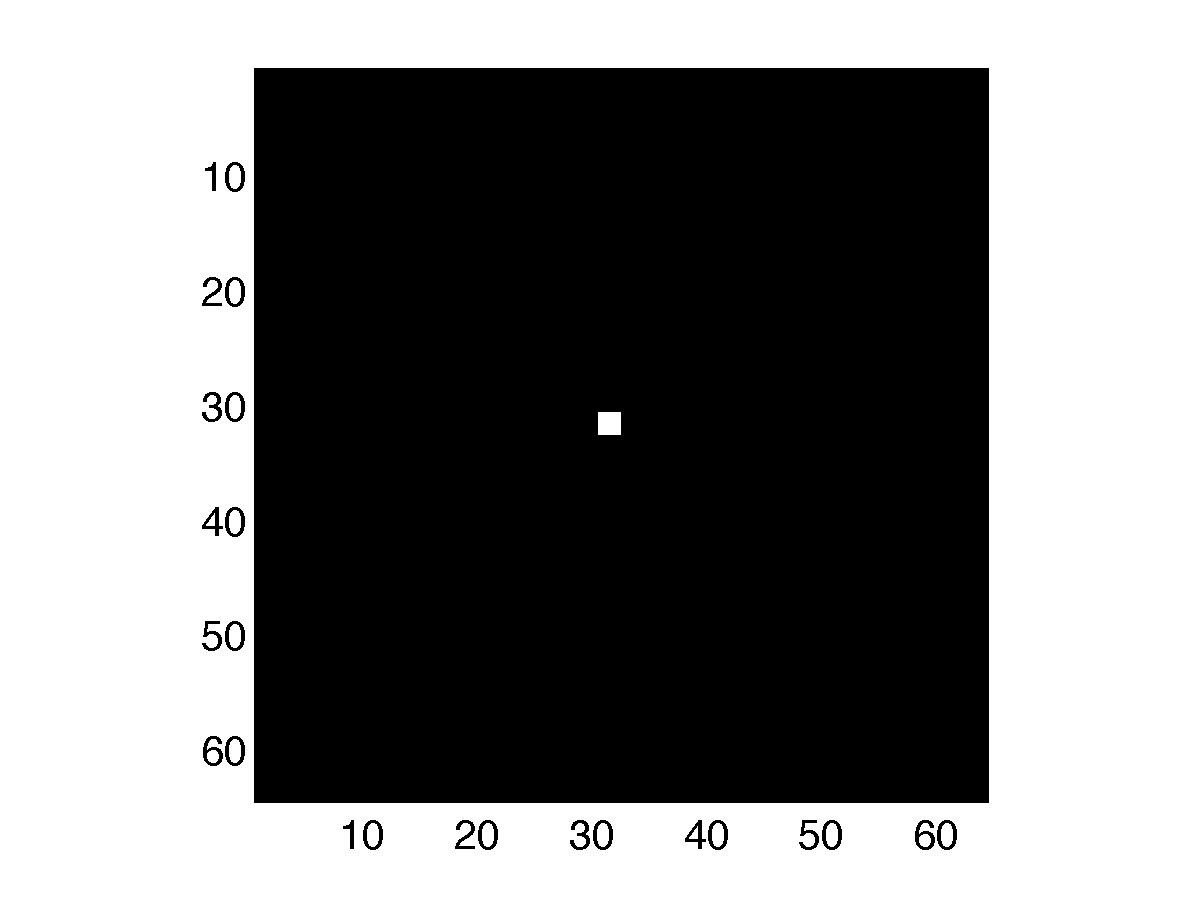}
\caption{\label{fig:T1} Each column of the matrix $T$ extracts an area of $2\times2$ pixels out of an image of ($64\times 64$) pixels.}
\end{minipage}
\\
\begin{minipage}[p]{0.45\linewidth}
\centering
    \includegraphics[width=1\linewidth]{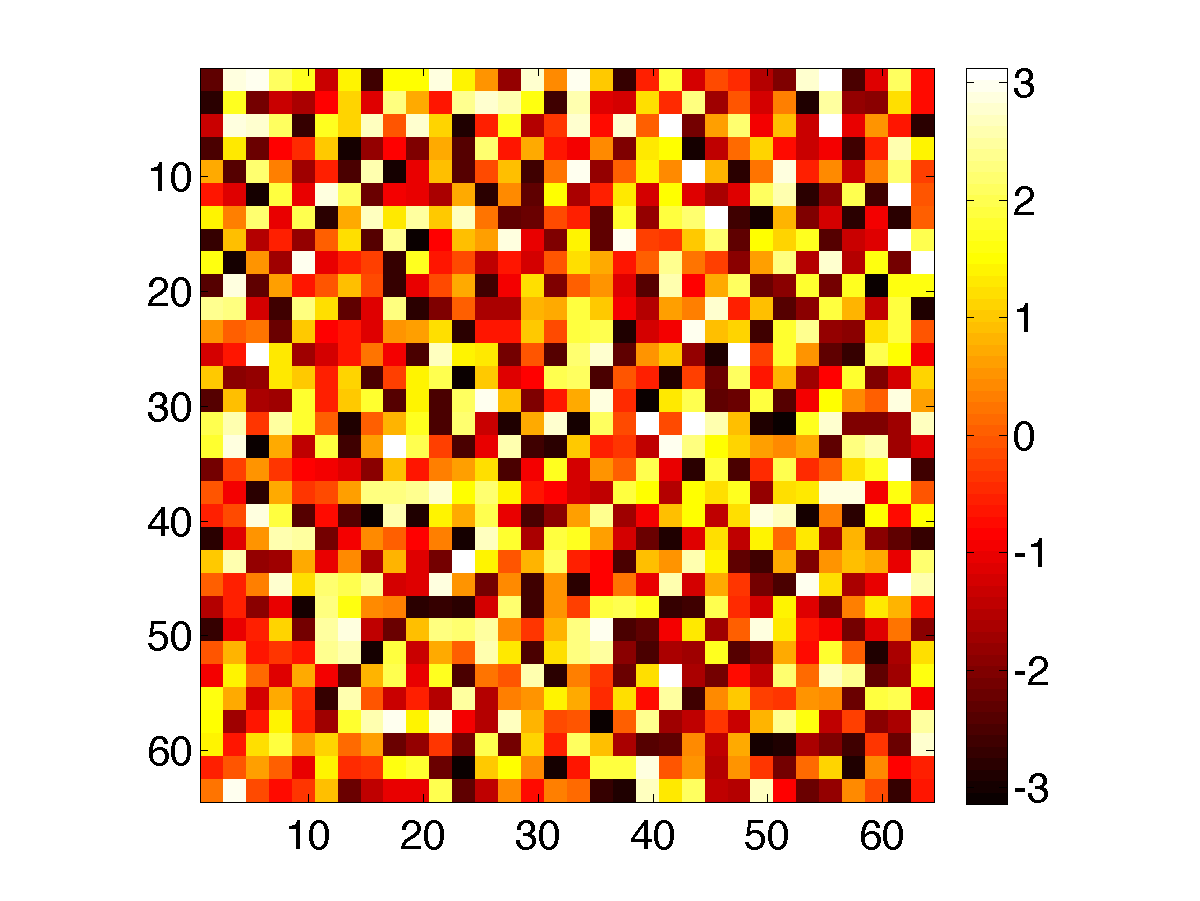}
\caption{\label{fig:phases}
Random perturbation: ($64\times 64$) phases
generated by $32\times 32$ random phases each spread over a bin of ($2\times 2$) pixels.
}
\end{minipage}
\hspace{0.05\linewidth}
\begin{minipage}[p]{0.45\linewidth}
\centering
    \includegraphics[width=1\linewidth]{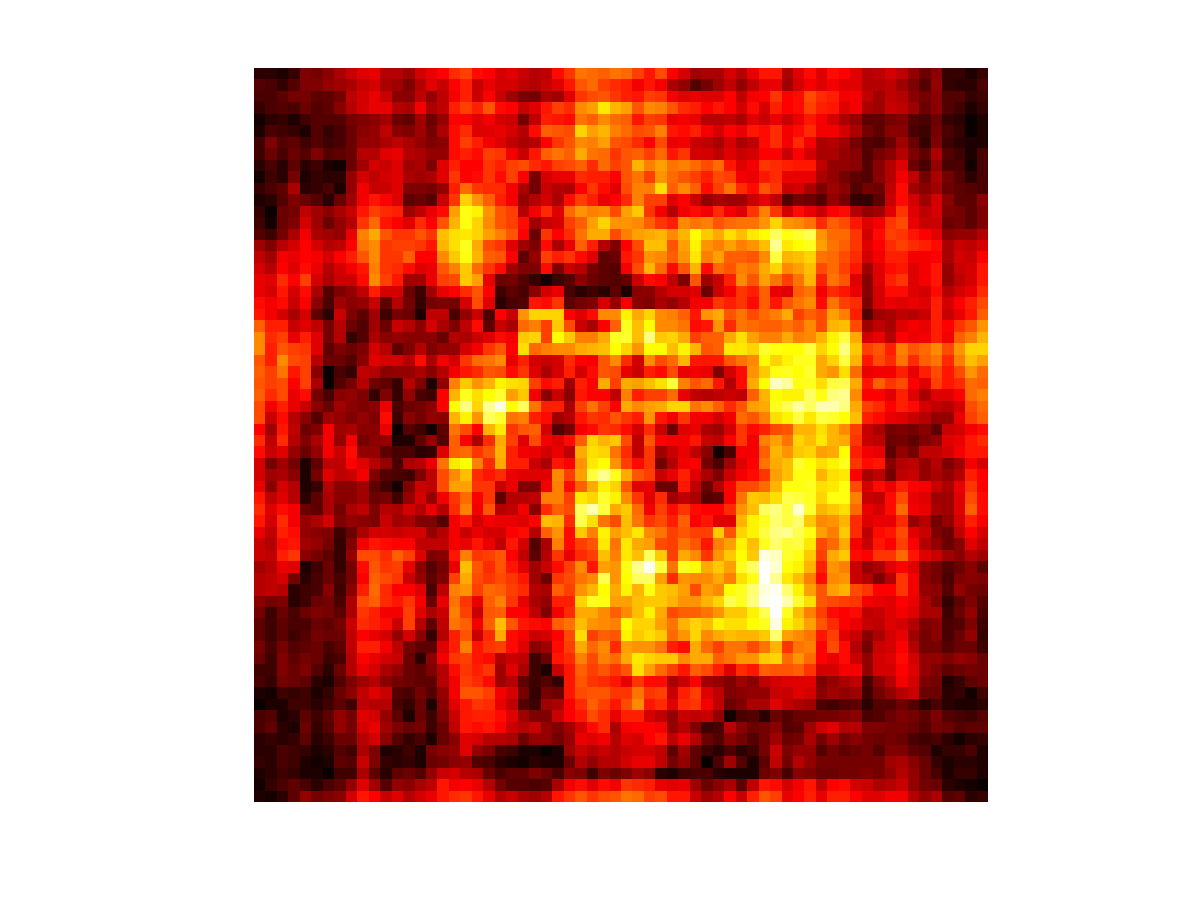}
\caption{\label{fig:gm} Image in real space ($\left |\bm F^\ast \bz \right|$) after random phase perturbation (Fig. \ref{fig:phases})}
\end{minipage}\\
\begin{minipage}[p]{0.45\linewidth}
\centering
    \includegraphics[width=1\linewidth]{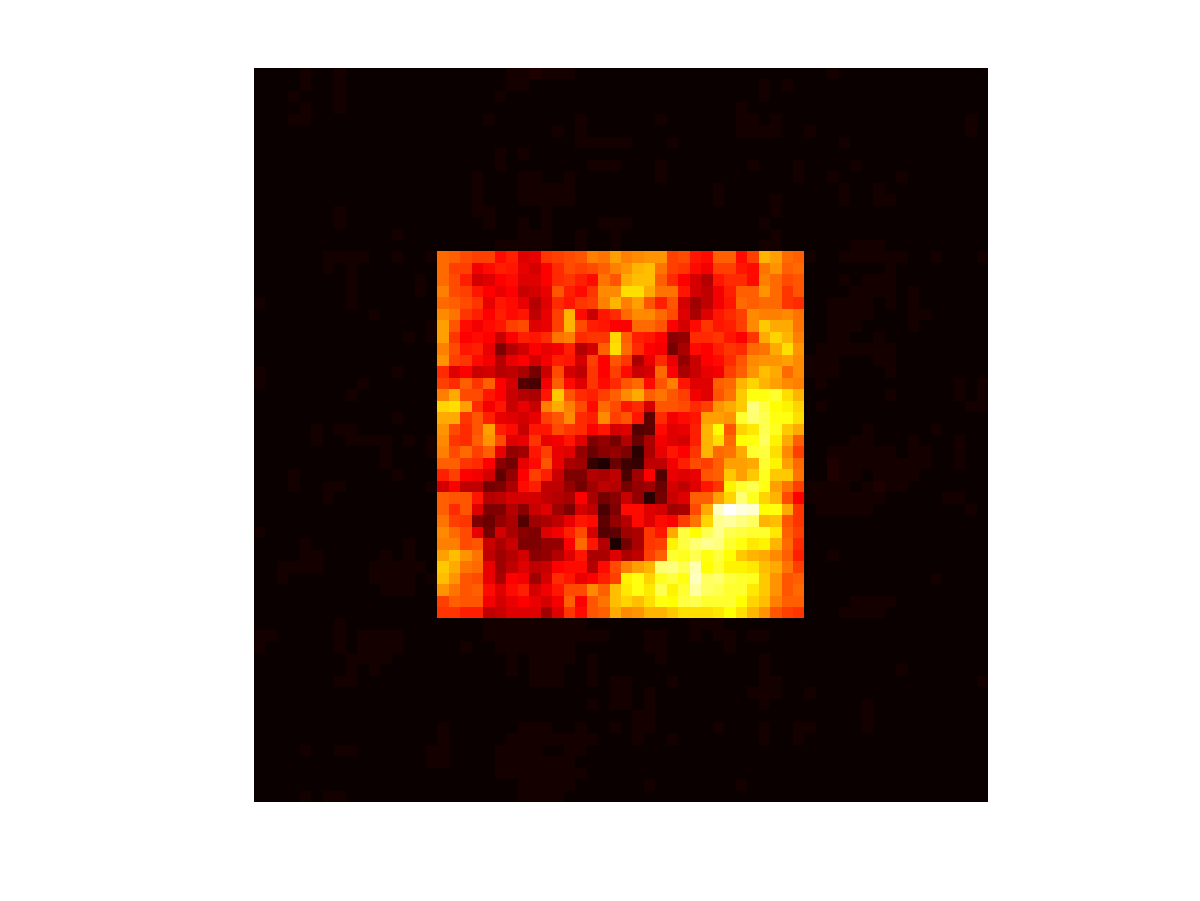}
\caption{\label{fig:gmAP} Image in real space after random phase perturbation 
(Fig. \ref{fig:phases}), using Eq. \ref{eq:synchronization} updates ( $\left |\bm F^\ast (P_{\bm a} P_S)^{1000}\bz \right|$).}
\end{minipage}
\hspace{0.05\linewidth}
\begin{minipage}[p]{0.45\linewidth}
\centering
    \includegraphics[width=1\linewidth]{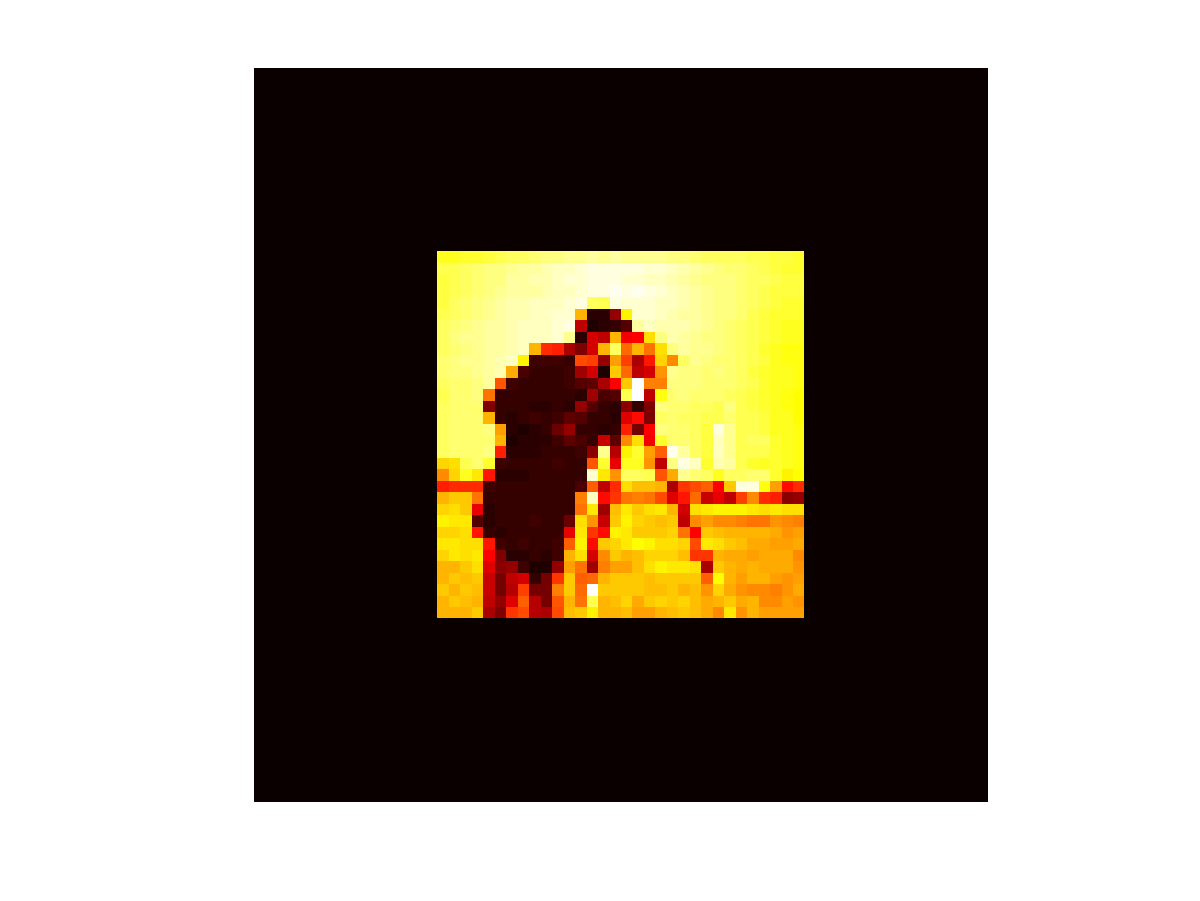}
\caption{\label{fig:gmw} Image in real space after one step of Eqs. (\ref{eq:augmented2},\ref{eq:augmented3})
 update.}
\end{minipage}
\end{figure*}

Here $\psi$ consists of the cameraman image of $32\times 32$ pixels,
embedded in a matrix of $64\times64$ pixels (Fig. \ref{fig:psi}).  The
``illumination matrix'' is the support of the object, $\Q=\Diag{S}$.
The support is 1 inside the $32\times 32$ box containing the image,
and 0 otherwise. $(\Q^\ast \Q)^{-1}$ is replaced by the pseudoinverse $\Q=\Q^\ast$. 
  The Fourier transform of $\psi$ was perturbed by
$32\times32$ randomly distributed phases (Fig. \ref{fig:phases}), each
multiplying a bin of $2\times2$ pixels (Fig. \ref{fig:T1}).  Upon
perturbation, the image in real space (Fig. \ref{fig:gm}) cannot be
distinguished.  Many iterations of Eq. \ref{eq:synchronization} or Eq.
\ref{eq:phasecut} cannot converge (Fig. \ref{fig:gmAP} showing Eq.
(\ref{eq:synchronization} ) updates), while 1 iteration of Eqs.
(\ref{eq:augmented2},\ref{eq:augmented3}) converges to the solution
(Fig. \ref{fig:gmw}).

\section{Conclusions}
I have shown that power iteration methods can recover phase
perturbations under special circumstances. If one is given the
relative phases between a small group of pixels (binned) and a random
perturbation of the phase between all the groups of pixels (the bins),
then the inverse block iteration can recover the solution in 1
iteration.  In \cite{augmented} it was observed that the inverse block
iteration improved convergence rate in large scale ptychographic
imaging. The inverse block iteration was also shown to recover
perturbations in the experimental geometry such as position errors and
intensity fluctuations. 
More work is needed to determine the optimal combination of Eqs.
(\ref{eq:synchronization},\ref{eq:phasecut},\ref{eq:reweight},\ref{eq:augmented2},\ref{eq:augmented3}),
and the properties of $T$,  in large scale phase retrieval problems.

I acknowledge usefull discussions with Jeff Donatelli of UC Berkeley.
This work was stimulated by the Phase Retrieval workshop at the Erwin
Schroedinger International Institute for Mathematical Physics (ESI)
organized by Karlheinz Gr\"ochenig and Thomas Strohmer.
This work was supported by the Laboratory Directed Research and
Development Program of Lawrence Berkeley National Laboratory under the
U.S. Department of Energy contract number DE-AC02-05CH11231.

\section*{Disclaimers}
This document was prepared as an account of work sponsored by the
United States Government. While this document is believed to contain
correct information, neither the United States Government nor any
agency thereof, nor the Regents of the University of California, nor
any of their employees, makes any warranty, express or implied, or
assumes any legal responsibility for the accuracy, completeness, or
usefulness of any information, apparatus, product, or process
disclosed, or represents that its use would not infringe privately
owned rights. Reference herein to any specific commercial product,
process, or service by its trade name, trademark, manufacturer, or
otherwise, does not necessarily constitute or imply its endorsement,
recommendation, or favoring by the United States Government or any
agency thereof, or the Regents of the University of California. The
views and opinions of authors expressed herein do not necessarily
state or reflect those of the United States Government or any agency
thereof or the Regents of the University of California.
\bibliography{ptyco}

\end{document}